\documentclass[10pt,twocolumn,hidelinks,preprint,nopreprintline]{elsarticle}
\pdfoutput=1
\usepackage{multirow}
\usepackage{tabularx}
\usepackage{graphicx}
\usepackage[table,xcdraw]{xcolor}
\usepackage[colorlinks=false]{hyperref}   
\usepackage{orcidlink}
\usepackage{cleveref} 
\usepackage{natbib} 
\usepackage{doi}

\newcommand{\mydimpar}[1]{\underline{\smash{#1}}}

\newcommand{\Myeg}{E.g., }
\newcommand{\myeg}{e.g., }
\newcommand{\myie}{i.e., }
\newcommand{\parencite}{\citep}
\newcommand{\textcite}{\citet}
\renewcommand{\cite}{\citep}

\definecolor{E}{rgb:Hsb}{210,0.67,0.35}
\definecolor{P}{rgb:Hsb}{210,0.67,0.59}
\definecolor{L}{rgb:Hsb}{210,0.56,0.78}
\definecolor{U}{rgb:Hsb}{210,0.27,0.88}

\definecolor{B}{rgb:Hsb}{210,0.67,0.59}
\definecolor{O}{rgb:Hsb}{210,0.56,0.78}
\definecolor{I}{rgb:Hsb}{210,0.27,0.88}

\definecolor{W}{rgb:Hsb}{210,0.56,0.78}
\definecolor{R}{rgb:Hsb}{210,0.27,0.88}

\definecolor{Iter}{rgb:Hsb}{210,0.56,0.78}

\definecolor{A}{rgb:Hsb}{210,0.67,0.59}
\definecolor{C}{rgb:Hsb}{210,0.56,0.78}
\definecolor{Con}{rgb:Hsb}{210,0.56,0.78}
\definecolor{D}{rgb:Hsb}{210,0.27,0.88}

\definecolor{calc}{rgb:Hsb}{210,0.27,0.88}
\definecolor{sing}{rgb:Hsb}{210,0.27,0.88}

\definecolor{form}{rgb:Hsb}{210,0.56,0.78}

\begin{document}

\Crefname{figure}{Figure}{Figures}
\crefname{figure}{figure}{figures}
\arrayrulecolor{lightgray}


\begin{frontmatter}

\title{Large Language Models as Software Components: \\ 
A Taxonomy for LLM-Integrated Applications 
}

\author{Irene Weber \orcidlink{0000-0003-2743-1698}} 

\address{
Kempten University of Applied Sciences, Germany \\
irene.weber@hs-kempten.de} 

\begin{abstract}
Large Language Models (LLMs) have become widely adopted recently. Research explores their use both as autonomous agents and as tools for software engineering. LLM-integrated applications, on the other hand, are software systems that leverage an LLM to perform tasks that would otherwise be impossible or require significant coding effort. While LLM-integrated application engineering is emerging as new discipline, its terminology, concepts and methods need to be established.
This study provides a taxonomy for LLM-integrated applications, offering a framework for analyzing and describing these systems. It also demonstrates various ways to utilize LLMs in applications, as well as options for implementing such integrations.

Following established methods, we analyze a sample of recent LLM-integrated applications to identify relevant dimensions. We evaluate the taxonomy by applying it to additional cases.
This review shows that applications integrate LLMs in numerous ways for various purposes. Frequently, they comprise multiple LLM integrations, which we term ``LLM components''. To gain a clear understanding of an application's architecture, we examine each LLM component separately. We identify thirteen dimensions along which to characterize an LLM component, including the LLM skills leveraged, the format of the output, and more. LLM-integrated applications are described as combinations of their LLM components. We suggest a concise representation using feature vectors for visualization.

The taxonomy is effective for describing LLM-integrated applications. It can contribute to theory building in the nascent field of LLM-integrated application engineering and aid in developing such systems. Researchers and practitioners explore numerous creative ways to leverage LLMs in applications. Though challenges persist, integrating LLMs may revolutionize the way software systems are built.

\end{abstract}


\end{frontmatter}


\section{Introduction}\label{sec:introduction}

Large Language Models (LLMs) have significantly impacted various sectors of economy and society \cite{raiaanReviewLargeLanguage2024}. Due to their proficiency in text understanding, creative work, communication, knowledge work, and code writing, they have been adopted in numerous fields, such as medicine, law, marketing, education, human resources, etc.

Public discussions often focus on the ethical aspects and societal consequences of these systems \cite{mcleanRisksAssociatedArtificial2023, mitchellDebatesNatureArtificial2024}. Meanwhile, research investigates Artificial General Intelligences and autonomous AI agents that can use services, data sources, and other tools, and collaborate to solve complex tasks \cite{chengExploringLargeLanguage2024,wangSurveyLargeLanguage2024,topsakal2023creating,handlerTaxonomyAutonomousLLMPowered2023}.
In addition, LLMs offer many opportunities to enhance software systems. They enable natural language interaction \cite{wangEnablingConversationalInteraction2023a}, automate complex tasks \cite{guanIntelligentVirtualAssistants2023}, and provide supportive collaboration, as seen with recent LLM-based assistant products often branded as ``copilots''\footnote{\Myeg \url{https://docs.github.com/en/copilot}, \\ \url{https://copilot.cloud.microsoft/en-us/copilot-excel}, \\ \url{https://www.salesforce.com/einsteincopilot}}.

This paper addresses the potential of LLMs for software development by integrating their capabilities as components into software systems. This contrasts with current software engineering research, which views LLMs as tools for software development rather than as software components \cite{fanLargeLanguageModels2023,HouLargeLanguageModels2023}, and with the  considerable body of research examining LLMs as autonomous agents within multiagent systems \cite{handlerTaxonomyAutonomousLLMPowered2023}.

Software systems that invoke an LLM and process its output are referred to as ``LLM-integrated applications'', ``LLM-integrated systems'', ``LLM-based applications'', etc. \cite{liuPromptInjectionAttack2023,evertzWhispersMachineConfidentiality2024,topsakal2023creating}.
LLMs are versatile, multipurpose tools capable of providing functionalities that would otherwise be unfeasible or require substantial development efforts \cite{fanRecommenderSystemsEra2023, jablonka14ExamplesHow2023}. By significantly expediting system development, they have the potential to revolutionize not only the way users interact with technology, but also the fundamental processes of software development.

LLM-integrated applications engineering is emerging as a research field. \Myeg \cite{chenSystemsEngineeringIssues2024} proposes LLM Systems Engineering (LLM-SE) as a novel discipline, and \cite{parninBuildingYourOwn2023, carterWeShippedAI2023, carterAllHardStuff2023} discuss experiences and challenges that developers of such systems encounter in practice. 

This study develops a taxonomy that provides a structured framework for categorizing and analyzing LLM-integrated applications across various domains. To develop and evaluate the taxonomy, we collected a sample of LLM-integrated applications, concentrating on technical and industrial domains. These applications showcase a broad range of opportunities to leverage LLMs, often integrating LLMs in multiple ways for distinct purposes.
In developing the taxonomy, we found that examining each of these integrations, termed ``LLM components'', separately is crucial for a clear understanding of an application's architecture.

The taxonomy adopts an original architectural perspective, focusing on how the application interacts with the LLM while abstracting from the specifics of application domains. 
For researchers, the taxonomy contributes to shape a common understanding and terminology, thus aiding theory building in this emerging domain \cite{kundischUpdateTaxonomyDesigners2022, schoormannExploringPurposesUsing2022, gregor2006nature}. For practitioners, the taxonomy provides inspiration for potential uses of LLMs in applications, presents design options, and helps identify challenges and approaches to address them.

\paragraph*{Objectives}
In this study, a taxonomy is understood as a set of dimensions divided into characteristics. The objective is to identify dimensions that are useful for categorizing the integration of LLMs in applications from an architectural perspective. To be most effective, the taxonomy should be easy to understand and apply, yet distinctive enough to uncover the essential aspects. Additionally, we aim to develop a visual representation tailored to the  taxonomy's intended purposes.

\paragraph*{Overview}
The following \cref{sec:background} provides background on LLMs and introduces
relevant concepts. \Cref{sec:related} presents an overview of related work. The
study design adheres to a \emph{Design Science Research} approach
\parencite{peffersDesignScienceResearch2007}. We apply established methods for taxonomy design
\parencite{nickersonMethodTaxonomyDevelopment2013,
rittelmeyerMorphologicalBoxAI2023} as described in \Cref{sec:method}. This
section also presents the sample of LLM-integrated applications used for this
study. The developed taxonomy is presented, demonstrated and formally evaluated
in \cref{sec:tax}. In \cref{sec:discussion}, we discuss its usability and
usefulness. \Cref{sec:conclusion} summarizes the contributions, addresses
limitations, and concludes.

\section{Large Language Models}\label{sec:background}

\subsection{Background}
State-of-the-art LLMs such as GPT-3.5, GPT-4, Llama, PALM2, etc., are
artificial neural networks consisting of neurons, \myie very simple processing
units, that are organized in layers and connected by weighted links. Training a
neural network means adapting these weights such that the neural network shows a
certain desired behavior. Specifically, an LLM is trained to predict the
likelihoods of pieces of text termed, \emph{tokens}, to occur as continuations of
a given text presented as input to the LLM. This input is referred to as \emph{prompt}. The
prompt combined with the produced output constitutes the \emph{context} of an
LLM. It may comprise more than 100k tokens in state-of-the-art
LLMs\footnote{\url{https://platform.openai.com/docs/models}}. Still, its length
is limited and determines the maximum size of prompts and outputs that an LLM is
capable of processing and generating at a time.

Training of an LLM optimizes its parameters such that its computed likelihoods
align with real text examples. The training data is a vast body of text snippets
extracted, processed, and curated from sources such as Wikipedia, Github code
repositories, common websites, books, or news archives. An LLM trained on massive
examples is termed a \emph{foundation model} or \emph{pre-trained model}. During training, an LLM not only learns to produce correct language but also absorbs and stores information and factual knowledge. However, it is well known that LLMs frequently pick up biases, leading to ethical problems. They may also produce factually incorrect outputs that sound plausible and convincing, termed \emph{hallucinations}.

Recent findings show that LLMs can be applied to a wide range of tasks by
appropriately formulating prompts. Different prompt patterns succeed in
different tasks. Basic approaches rely on instructing the LLM to solve a task
described or explained in the prompt. In \textit{few-shot} prompting (also known
as \textit{few-shot} learning), the prompt is augmented with example
input-output pairs illustrating how to solve the task, \myeg the requested
output format. The number of examples can vary. Prompting with one example is called \textit{one-shot} prompting, while prompting without any examples is called \textit{zero-shot} prompting. \textit{One-shot} and
\textit{few-shot} prompting fall under the broader category of
\textit{in-context learning}. Prompt patterns such as \textit{chain-of-thought}
and \textit{thinking-aloud} aim to elicit advanced reasoning
capabilities from LLMs.

As effective prompts are crucial for unlocking the diverse capabilities of an
LLM, the discipline of prompt engineering is evolving, focusing on the systematic design and management of prompts \cite{whitePromptPatternCatalog2023a,
chenUnleashingPotentialPrompt2023,strobeltInteractiveVisualPrompt2022,liuPretrainPromptPredict2023}.

\subsection{Definitions}\label{sec:definitions}

Invoking an LLM results in an input-processing-output sequence: Upon receiving a
prompt, the LLM processes it and generates an output. We refer to an individual sequence of
input-processing-output performed by the LLM as \emph{LLM invocation}, and 
define an \emph{LLM-integrated application} as a system in which the software
generates the prompt for the LLM and processes its output. The concept of an application is broad, encompassing service-oriented architectures and systems with components loosely coupled via API calls.

Given an LLM's versatility, an application can utilize it for different tasks, each demanding a specific approach to create the prompt and handle the result.
This paper defines a particular software component that accomplishes this as an \emph{LLM-based software component} or, simply, \emph{LLM component}. An LLM-integrated application can comprise several LLM
components. The study develops a taxonomy for LLM components. LLM-integrated applications are described as combinations of their LLM components. 

\section{Related Work}\label{sec:related}

With the recent progress in generative AI and LLMs, the interest in these
techniques has increased, and numerous surveys have been published,
providing an extensive overview of
technical aspects of LLMs \parencite{zhaoSurveyLargeLanguage2023}, reviewing 
 LLMs as tools for software engineering  \parencite{HouLargeLanguageModels2023}, and
 discussing the technical
challenges of applying LLMs across various fields \parencite{kaddourChallengesApplicationsLarge2023}. Further studies address
the regulatory and ethical
aspects of Generative AI and ChatGPT, with a particular focus on AI-human
collaboration \parencite{nahGenerativeAIChatGPT2023}, and Augmented
Language Models (ALMs), which are LLMs that enhance their capabilities by
querying tools such as APIs, databases, and web search engines  \parencite{mialonAugmentedLanguageModels2023}. 

Taxomonies related to LLMs include a taxonomy for prompts designed to solve complex tasks \parencite{santuTELeRGeneralTaxonomy2023} and a taxonomy of methods for cost-effectively invoking a remote LLM \cite{wangSurveyEffectiveInvocation2024}.
A comparative analysis of studies on applications of ChatGPT is provided by \parencite{koubaaExploringChatGPTCapabilities2023}, whereas LLMs are compared based on their application domains and the tasks they solve in \parencite{hadiLargeLanguageModels2023}.
Most closely related to the taxonomy developed here is a  taxonomy for LLM-powered multiagent architectures \parencite{handlerTaxonomyAutonomousLLMPowered2023} which focuses on autonomous agents with less technical detail.
Taxonomies of applications of AI in enterprises \parencite{rittelmeyerMorphologicalBoxAI2023} and applications of generative AI, including but not limited to LLMs \parencite{strobelExploringGenerativeArtificial2024}, are developed using methods similar to those in our study.

Several taxonomies in the field of conversational agents and task-oriented dialog (TOD) systems address  system architecture
\parencite{adamopoulouOverviewChatbotTechnology2020,
motgerSoftwareBasedDialogueSystems2022,
colabianchiHumantechnologyIntegrationIndustrial2023,baezChatbotIntegrationFew2020}.
However, they omit detailed coverage of the integration of generative language models.

\section{Methods}\label{sec:method}

We constructed the taxonomy following established guidelines
\parencite{nickersonMethodTaxonomyDevelopment2013,
rittelmeyerMorphologicalBoxAI2023, kundischUpdateTaxonomyDesigners2022}, drawing from a sample of LLM-integrated applications. These applications are detailed in
\cref{sec:instancecollect}.

\subsection{Development}
\paragraph{Taxonomy}
We derived an initial taxonomy from the standard architecture of
conversational assistants described in
\parencite{baezChatbotIntegrationFew2020}, guided by the idea that conversational assistants are essentially ``chatbots with
tools'', \myie language-operated user interfaces that interact with external
systems. This approach proved unsuccessful.
The  second version was based on the classical three-tier software architecture, and then extended over several development cycles. By repeatedly applying the evolving taxonomy to the example instances, we identified dimensions and characteristics using an ``empirical-to-conceptual'' approach. When new dimensions emerged, additional 
characteristics were derived in a ``conceptual-to-empirical'' manner. 
After five major refinement cycles, the set of dimensions and characteristics 
solidified. In the subsequent evaluation phase, we applied the taxonomy to a new
set of example instances that were not considered while constructing the
taxonomy. As the dimensions and characteristics remained stable, the taxonomy
was considered complete. In the final phase, we refined the wording and visual format of the taxonomy.

\paragraph{Visualization} Developing a taxonomy involves creating a
representation that effectively supports its intended purpose
\cite{kundischUpdateTaxonomyDesigners2022}. Taxonomies can be represented in
various formats, with morphological boxes 
\cite{szopinskiCriteriaPreludeGuiding2020,szopinskivisualize2020} or 
radar charts \cite{handlerTaxonomyAutonomousLLMPowered2023} being well-established approaches.
We evaluated morphological boxes, because they  effectively position categorized instances within the design space. However, we found that they make it difficult to perceive a group of categorized instances as a whole since they occupy a large display area. This drawback is
significant for our purposes, as LLM-integrated applications often comprise
multiple LLM components. Therefore, we developed a more condensed visualization of the taxonomy based on feature vectors.

\paragraph{Example instances} \label{sec:instancecollect}

We searched for instances of LLM-integrated applications for taxonomy development
that should meet the following criteria:

\begin{itemize}
\item The application aims for real-world use rather than focusing on research only (such as testbeds for experiments or proofs-of-concept). It demonstrates efforts towards practical usability and addresses challenges encountered in real-world scenarios.
\item The application's architecture, particularly its LLM components, is described in sufficient detail for analysis.
\item The sample of instances covers a diverse range of architectures.
\item The example instances are situated within industrial or technical domains,
as we aim to focus on LLM-integrated applications beyond well-known fields like law, medicine, marketing, human resources, and education.
\end{itemize}

The search revealed a predominance of theoretical
research on LLM-integrated applications while papers focusing on practically
applied systems were scarce. Searching non-scientific websites uncovered 
commercially advertised AI-powered applications, but their internal workings were typically undisclosed, and reliable evaluations were lacking.
Furthermore, the heterogeneous terminology and concepts in this emerging field
make a comprehensive formal literature search unfeasible. Instead, by repeatedly searching Google Scholar and non-scientific websites using terms ``LLM-integrated
applications'', ``LLM-powered applications'', ``LLM-enhanced system'', ``LLM'' and
``tools'', along similar variants, we selected six suitable 
instances. Some of them
integrate LLMs in multiple ways, totaling eleven distinct LLM components.

For a thorough evaluation, we selected new instances using relaxed criteria, including those intended for research. Additionally, we included a real-world example lacking explicit documentation to broaden the diversity of our sample and assess the taxonomy's coverage. Within the five selected instances, we identified ten LLM components.

\subsection{Sample of LLM-integrated applications} \label{sec:cases}
Table~\ref{tab:cases} gives an overview of the sample. Names of applications and LLM components are uniformly written as one CamelCase word and typeset in small caps, deviating from the format chosen by the respective authors.

\begin{table*}[t]
 \caption{Example instances selected for development (top 6) and evaluation (bottom 5) }\label{tab:cases}

\medskip

 \begin{tabularx}{\textwidth}{llX}   
\hline
   Application & References & LLM components \\
 \hline 
   \textsc{Honeycomb} & \cite{carterAllHardStuff2023, carterWeShippedAI2023} & \textsc{QueryAssistant} \\
   \textsc{LowCode} & \cite{caiLowcodeLLMVisual2023},\cite{maoLowCodeLLM2023} & \textsc{Planning}, \textsc{Executing} \\
   \textsc{MyCrunchGpt}& \cite{kumarMyCrunchGPTLLMAssisted2023} & \textsc{DesignAssistant}, \textsc{SettingsEditor}, 
   \mbox{\textsc{DomainExpert}} \\
   \textsc{MatrixProduction}& \cite{XiaAutonomousSystemFlexible2023} & \textsc{Manager}, \textsc{Operator} \\
   \textsc{Workplace\-Robot} & \cite{meesGroundingLanguageVisual2023} & \textsc{TaskPlanning} \\
   \textsc{AutoDroid} & \cite{wenEmpoweringLLMUse2023} & \textsc{TaskExecutor}, \textsc{MemoryGenerator} \\
   \hline
   \hline
   \textsc{ProgPrompt} & \cite{singhProgPromptGeneratingSituated2023} & \textsc{ActionPlanning}, \textsc{ScenarioFeedback} \\
   \textsc{FactoryAssistants} & \cite{kernanfreireHarnessingLargeLanguage2023} & \textsc{QuestionAnswering} \\
   \textsc{Sgp\-Tod} & \cite{zhangSGPTODBuildingTask2023} & \textsc{DstPrompter}, \textsc{PolicyPrompter} \\
   \textsc{TruckPlatoon} & \cite{dezarzaLLMAdaptivePID2023} & \textsc{Reporting} \\
   \textsc{ExcelCopilot} & \cite{fortinMicrosoftCopilotExcel2024,parninBuildingYourOwn2023} & \textsc{ActionExecutor}, \textsc{Advisor}, 
   \mbox{\textsc{IntentDetector}}, \textsc{Explainer} \\
   \hline 
 \end{tabularx}
\renewcommand{\arraystretch}{1}
 \end{table*}

\paragraph{\textsc{Honeycomb}}
\label{Honeycomb-query-assistant} 
\textsc{Honeycomb} is an observability platform collecting data from software applications in distributed environments for monitoring. Users define queries to retrieve information about the observed software systems through \textsc{Honeycomb}’s Query Builder UI.
The recently added LLM-based \textsc{QueryAssistant} allows users to articulate inquiries in plain English, such as ``slow endpoints by status code'' or ``which service has the highest latency?'' The \textsc{QueryAssistant} converts these into queries in \textsc{Honeycomb}'s format, which users can execute and manually refine	
 \cite{carterAllHardStuff2023, carterWeShippedAI2023}.

\paragraph{\textsc{LowCode}}\label{low-code-llm-visual-programming-over-llms-x}

\textsc{LowCode} is a web-based application consisting of a prompt-definition section and a dialogue section. The prompt-definition section supports the design of prompts for complex tasks, such as composing extensive essays, writing resumes for job applications or acting as a hotel service chatbot \parencite{caiLowcodeLLMVisual2023}. In the dialogue section, users converse with an LLM to complete the complex task based on the defined prompt.

\textsc{LowCode} comprises two LLM components termed \textsc{Planning} and \textsc{Executing}. \textsc{Planning} operates in the prompt-definition section, where a user roughly describes a complex task, and \textsc{Planning}
designs a workflow for solving it. The prompt-definition section offers a low-code development environment where the LLM-generated workflow is visualized as a graphical flowchart, allowing a user to edit and adjust the logic of the flow and the contents of its steps. For instance, in essay-writing scenarios, this involves inserting additional sections, rearranging sections, and refining the contents of sections. 
Once approved by the user, \textsc{LowCode} translates the modified workflow back into natural language and incorporates it into a prompt for \textsc{Executing}. In the dialogue section, users converse in interactive, multi-turn dialogues with \textsc{Executing}. As defined in the prompt, it acts as an assistant for tasks such as writing an essay or resume, or as a hotel service chatbot. 
While the idea of the LLM planning a workflow might suggest using the LLM for application control, \textsc{LowCode Planning} actually serves as a prompt generator that supports developing prompts for complex tasks.

\paragraph{\textsc{MyCrunchGpt}}\label{MyCrunchGpt}
 \textsc{MyCrunchGpt} acts as an expert system within the engineering domain, specifically for airfoil design and calculations in fluid mechanics. These tasks require complex workflows comprising several steps such as preparing data, parameterizing tools, and evaluating results, using various software systems and tools. The aim of \textsc{MyCrunchGpt} is to facilitate the definition of these workflows and automate their execution \cite{kumarMyCrunchGPTLLMAssisted2023}.

\textsc{MyCrunchGpt} offers a web interface featuring a dialogue window for inputting commands in plain English, along with separate windows displaying the output and results of software tools invoked by \textsc{MyCrunchGpt} in the backend.
\textsc{MyCrunchGpt} relies on predefined workflows, not supporting deviations or cycles. By appending a specific instruction to the dialogue history in the prompt for each step of the workflow, it uses the LLM as a smart parser to extract parameters for APIs and backend tools from user input. APIs and tools are called in the predefined order \cite[p.~56]{kumarMyCrunchGPTLLMAssisted2023}.

\textsc{MyCrunchGpt} is still in development. The paper \parencite{kumarMyCrunchGPTLLMAssisted2023} explains the domain as well as the integration of the LLM, but does not fully detail the implementation of the latter. Still, \textsc{MyCrunchGpt} illustrates innovative applications of an LLM in a technical domain. We categorize three LLM components solving tasks within \textsc{MyCrunchGpt}: a \textsc{DesignAssistant} guiding users through workflows and requesting parameters for function and API calls; a \textsc{SettingsEditor} updating a JSON file with settings for a backend software tool; and a \textsc{DomainExpert} which helps evaluating results by comparing them to related results, \myeg existing airfoil designs, which it derives from its trained knowledge.

\paragraph{\textsc{MatrixProduction}}

\textsc{MatrixProduction} employs an LLM for controlling a matrix production system \parencite{XiaAutonomousSystemFlexible2023}. 
While in a classical line production setup,  workstations are arranged linearly and the manufacturing steps follow a fixed sequence, matrix production is oriented towards greater flexibility. Autonomous transport vehicles carry materials and intermediate 
products to workstations, termed automation modules, each offering a spectrum of manufacturing skills that it can contribute to the production process.
Compared to line production, matrix production is highly adaptable and can manufacture a variety of personalized products with full automation. This requires intelligent production management to (a) create workplans that orchestrate and schedule the automation modules’ skills, and (b) program 
the involved automation modules such that they execute the required processing steps. 

\textsc{MatrixProduction} incorporates two LLM components: \textsc{Manager} creates workplans as sequences of skills (a), while \textsc{Operator} generates programs for the involved automation modules (b).

\textsc{MatrixProduction} prompts \textsc{Manager} and \textsc{Operator} to
provide textual explanations in addition to the required sequences of skills or
automation module programs. The LLM output is processed by a parser before
being used to control the physical systems. \textsc{Manager} relies on built-in
production-specific knowledge of the LLM such as ``a hole is produced by
drilling''.

Noteworthy in this approach is its tight integration into the system landscape of Industry 4.0. The \textit{few-shot} \textsc{Manager} and \textsc{Operator} prompts are generated automatically using \textit{Asset Administration Shells}, which are standardized, technology-independent data repositories storing digital twins of manufacturing assets for use in Industry 4.0 \cite{bader2022details}. 

\paragraph{\textsc{Workplace\-Robot}}\label{sec-robot-control} An experimental
robot system is enhanced with LLM-based task planning in
\parencite{meesGroundingLanguageVisual2023}. The robot operates in a workplace
environment featuring a desk and several objects. It has previously been trained to
execute basic operations expressed in natural language such as
``open the drawer'' or ``take the pink object and place it in the drawer''.
LLM-based task planning enables the robot to perform more complex orders like
``tidy up the work area and turn off all the lights''. To this end, an LLM is
prompted to generate a sequence of basic operations that accomplish
the complex order. 

Although the robot expects operations phrased in natural
language, the LLM is prompted with a Python coding task. For instance, the basic operation ``turn on the green light'' corresponds to a Python command
\texttt{push\_button('green')}. The prompt for the LLM includes several examples
each consisting of a description of an environment state, a complex order
formatted as a comment, and a sequence of Python robot commands that accomplish
the complex order. When invoking the LLM to generate the Python program for a
new order, the prompt is augmented with a description of the environment's
current state and the new order as a comment. 

The Python code produced by the LLM
is translated back to a sequence of basic operations in natural
language. When the robot executes these operations, there is no feedback
about successful completion. Rather, the system assumes that all basic
operations require a fixed number of timesteps to complete.

\paragraph{\textsc{AutoDroid}} The goal of mobile task automation is hands-free user interaction for smartphones through voice commands. \textsc{AutoDroid} is a voice control system for smartphones that can automatically execute complex orders such as ``remind me to do laundry on May 11th'' or ``delete the last photo I took'' \cite{wenEmpoweringLLMUse2023, wenMobileLLMAutoDroid2024}.

Such complex orders are fulfilled by performing sequences of basic operations in an Android app, such as ``scroll down, then press button x'' in the calendar app.
\textsc{AutoDroid} employs an LLM component \textsc{TaskExecutor} to plan these
 sequences of operations. The challenge is that the next operation to execute depends on the current state of the Android app which continuously changes as the app is operated. \textsc{AutoDroid} solves this by invoking the
\textsc{TaskExecutor} repeatedly after each app operation with the prompt
comprising the updated state of the Graphical User Interface (GUI) along with the user's complex order. 

Before executing irrevocable operations, such as permanently deleting data or calling a contact, \textsc{AutoDroid} prompts the user to confirm or adjust the operation. \textsc{TaskExecutor} is instructed to include a ``confirmation needed'' hint in its output for such operations.

The prompt for \textsc{TaskExecutor} comprises an extract from a knowledge base which is built automatically in an offline learning phase as follows: In a first
step, a ``UI Automator'' (which is not an LLM component) automatically and
randomly operates the GUI elements of an Android app to generate a UI Transition
Graph (UTG). The UTG has GUI states as nodes and the possible transitions
between GUI states as edges. As next steps, \textsc{AutoDroid} invokes two LLM
components referred to as \textsc{MemoryGenerator}s to analyze the UTG.

The first \textsc{MemoryGenerator} is prompted
repeatedly for each GUI state in the UTG. Its task is to explain the
functionality of the GUI elements. Besides instructions and examples of the
table format desired as output, its prompt includes an HTML representation of
the GUI state, the GUI actions preceding this state, and the GUI element
operated next. Its output consists of tuples explaining the functionality of a
GUI element by naming the derived functionality (\myeg ``delete all the events
in the calendar app'') and the GUI states and GUI element
actions involved. Similarly, the second \textsc{MemoryGenerator} is prompted to output a table
listing GUI states and explanations of their functions. These tables constitute
\textsc{AutoDroid}'s knowledge base.

\paragraph{\textsc{ProgPrompt}} \textsc{ProgPrompt}
\cite{singhProgPromptGeneratingSituated2023} is an approach to LLM-based robot
task planning similar to \textsc{Workplace\-Robot}. Its robot is controlled by
Python code and works in a real and a simulated household environment. 

\textsc{ProgPrompt} comprises two LLM components. \textsc{ActionPlanning}
generates Python scripts for tasks such as ``microwave salmon'' using basic
operations like \texttt{grab('salmon')}, \texttt{open('microwave')}, and
\texttt{putin('salmon', 'microwave')}, notably without considering the
current state of the environment. To establish a feedback loop with the environment, \textsc{ActionPlanning} adds
\texttt{assert} statements. These statements verify the preconditions of basic
operations and trigger remedial actions when preconditions are not met.
For instance, a script for ``microwave salmon'' comprises the following
code fragment:

{\footnotesize\begin{verbatim}
   if assert('microwave' is 'opened')
      else:  open('microwave')
   putin('salmon', 'microwave')
\end{verbatim} }

When operating in the simulated environment, \textsc{ProgPrompt} can verify an
\texttt{assert} statement through its second LLM component, \textsc{Scenario\-Feedback}. Prompted with the
current state of the environment and the \texttt{assert} statement,
\textsc{Scenario\-Feedback} evaluates it and outputs \texttt{True} or
\texttt{False}.

\paragraph{\textsc{FactoryAssistants}} \textsc{FactoryAssistants} advise workers
on troubleshooting production line issues in two manufacturing domains:
detergent production and textile production
\parencite{kernanfreireHarnessingLargeLanguage2023}. The assistants leverage
domain knowledge from FAQs and documented problem cases to answer user queries.
The required domain knowledge is provided as a part of the
prompt.

\paragraph{\textsc{Sgp\-Tod}} \label{sec:SgpTod} \label{sec:intent-expl}
\textsc{Sgp\-Tod} employs an LLM to implement a chatbot, specifically, a
task-oriented dialogue (TOD) system \cite{zhangSGPTODBuildingTask2023}. TOD
systems are also known as conversational assistants. In contrast to open-domain
dialogue (ODD) systems, which engage users in goalless conversations, they are
designed for assisting users in specific tasks. 

In general, TOD systems require the following components
\cite{baezChatbotIntegrationFew2020}: Natural Language Understanding (NLU), analyzing the user's input to
classify intents and extract entities;  Dialogue
Management (DM) for deciding on a system action that is appropriate in a given
dialogue state (\myeg ask for more information or invoke a hotel booking
service); and Natural Language Generation (NLG) for producing a response that
the TOD system can present to the user. Intent classification, also known as intent detection, matches free-text user input to one of several tasks a TOD system can perform (\myeg book a hotel). Entity
extraction isolates situational values, called entities, from the user
input (\myeg the town and the date of the hotel booking). The TOD system may require several dialogue turns to elicit all necessary entities from
the user. In TOD research, the system’s internal representation of the
user's intentions and the entity values is commonly referred to as its ``belief state''.
For example, in the restaurant search domain, the belief state may include
attribute-value pairs like \texttt{cuisine:Indian} and
\texttt{pricerange:medium.}

\textsc{Sgp\-Tod} is a multi-domain TOD system, concurrently handling multiple
task domains found in standard TOD evaluation datasets, such as recommending
restaurants or finding taxis. Similar to other experimental TOD systems \cite{hudecekAreLargeLanguage2023}, \textsc{Sgp\-Tod} accesses a database that stores information from the task domains, such as available hotels and restaurants.

\textsc{Sgp\-Tod} comprises two LLM components, called \textsc{DstPrompter} and
\textsc{PolicyPrompter}, that are both invoked in every dialogue turn between
\textsc{Sgp\-Tod} and the user. The \textsc{DstPrompter} handles the NLU aspect,
analyzing the user’s input and populating the system’s belief state. It outputs
is an SQL query suited to extract the database entries that match the current belief
state. Upon retrieving the database entries, \textsc{Sgp\-Tod} invokes its
\textsc{PolicyPrompter} which covers both DM and NLG. Prompted with the dialogue
history and the database entries retrieved, it produces a two-part output: a natural language response for NLG and a system action for DM.

\paragraph{\textsc{TruckPlatoon}} The concept of truck platooning means that trucks travel
closely together for better fuel efficiency and traffic
flow. \textsc{TruckPlatoon} comprises an algorithmic control loop which
autonomously maintains a consistent distance between trucks. It invokes an LLM to
generate natural-language reports on the platoon’s performance and stability
from measurements tracked by the control algorithm, providing easily
understandable information for engineers involved in monitoring and optimizing
the truck platooning system.

\paragraph{\textsc{ExcelCopilot}} \label{sec:intent-expl2} 
\textsc{ExcelCopilot} is an example of a recent trend where software companies integrate LLM-based assistants, often termed ``copilots'', into their products \cite{parninBuildingYourOwn2023}. These copilots not only provide textual guidance but also perform actions within the software environment,
constituting a distinctive type of LLM-integrated application. We chose
\textsc{ExcelCopilot} as an example for evaluating our taxonomy. Since its implementation is undisclosed, we infer its architecture from indirect sources, including a screencast and a report on insights and experiences from copilot developers \cite{fortinMicrosoftCopilotExcel2024,parninBuildingYourOwn2023}. This inferred architecture may deviate from the actual implementation.

\textsc{ExcelCopilot} is accessible in a task bar alongside the Excel worksheet. It features buttons with context-dependent suggestions of actions and a text box for users to type in commands in natural language.
\textsc{ExcelCopilot} only works with data tables, so its initial suggestion is to convert the active worksheet's data into a data table. Copilot functions activate when a data table or part of it is selected.
It then presents buttons for four top-level tasks: ``add formula columns'',
``highlight'', ``sort and filter'', and ``analyze''. The ``analyze'' button triggers the
copilot to display more buttons, \myeg one that generates a pivot chart from the
selected data. \textsc{ExcelCopilot} can also add a formula column to the data
table and explain the formula in plain language.

When a user inputs a free-text command, \textsc{ExcelCopilot} may communicate
its inability to fulfill it. This constantly occurs with commands
requiring multiple steps, indicating that \textsc{ExcelCopilot} lacks a
planning LLM component as seen in, for example, \textsc{MatrixProduction}.
This observation, along with its mention in \cite{parninBuildingYourOwn2023}, suggests that \textsc{ExcelCopilot} employs an intent detection-skill routing architecture. This architecture includes an LLM component that maps free-text user commands to potential intents and then delegates to other LLM components tasked with generating actions to fulfill those intents.
Accordingly, \textsc{ExcelCopilot}
comprises several types of LLM components:

\begin{itemize}
 \item Several distinct \textsc{Action Executor}s generate code for specific application actions, such as creating a pivot table, designing a worksheet formula, inserting a diagram, and so on.
 \item An \textsc{Advisor} suggests meaningful next actions. Its outputs serve
 to derive button captions and prompts for \textsc{ActionExecutor}s.	
 \item When a user inputs a free-text command, the \textsc{IntentDetector} is invoked to determine and trigger a suitable \textsc{ActionExecutor}. The \textsc{IntentDetector} communicates its actions to users and informs them when it cannot devise a suitable action.
 \item 	The \textsc{Explainer} generates natural language explanations of formulae designed by \textsc{ExcelCopilot}. It is unclear whether under the hood, the \textsc{ActionExecutor} is generating both the formula and the explanation, or if two separate LLM components are being invoked. We assume the latter, \myie that a separate \textsc{Explainer} LLM component exists.
\end{itemize}
While users interact repeatedly with \textsc{ExcelCopilot}, each interaction
adheres to a single-turn pattern, with the user providing a command and
\textsc{ExcelCopilot} executing it \cite{parninBuildingYourOwn2023}.

 \section{A Taxonomy for LLM Components and LLM-Integrated Applications }\label{sec:tax}

When developing the taxonomy, it emerged that analyzing an LLM-integrated application should begin with identifying and describing its distinct LLM components. Analyzing each LLM component separately helps capture details and provides a clear understanding of how the application utilizes LLM capabilities. The LLM-integrated application
can then be described as a combination of the LLM components it employs. 

\begin{table*}[hbt]
 \caption{Dimensions and characteristics of the taxonomy. Codes of characteristics are printed in uppercase. ``Meta'' means ``metadimension''. ``MuEx'' means ``mutual exclusiveness''. } 
  \label{tab:dimover}
\medskip

\arrayrulecolor{lightgray}
  \begin{tabularx}{\textwidth}{XXlX}
\hline
     Meta & Dimension & Characteristics & MuEx \\
\hline  
     \textit{Invocation} & \textit{Interaction} & \textit{\textbf{A}pp}, \textit{\textbf{C}ommand}, \textit{\textbf{D}ialog} & enforced \\
     & \textit{Frequency} & \textit{\textbf{S}ingle}, \textit{\textbf{I}terative} & yes   \\
     \hline
     \textit{Function}  & \textit{Logic} & \textit{c\textbf{A}lculate}, \textit{\textbf{C}ontrol} & yes \\
    & \textit{UI} & \textit{none}, \textit{\textbf{I}nput}, \textit{\textbf{O}utput}, \textit{\textbf{B}oth} & yes \\
       & \textit{Data} & \textit{none}, \textit{\textbf{R}ead}, \textit{\textbf{W}rite}, \textit{\textbf{B}oth} & yes \\
     \hline
     \textit{Prompt} & \textit{Instruction} & \textit{none}, \textit{\textbf{U}ser}, \textit{\textbf{L}LM}, \textit{\textbf{P}rogram} & enforced\\ 
     & \textit{State} & \textit{none}, \textit{\textbf{U}ser}, \textit{\textbf{L}LM}, \textit{\textbf{P}rogram} & enforced \\ 
     & \textit{Task} & \textit{none}, \textit{\textbf{U}ser}, \textit{\textbf{L}LM}, \textit{\textbf{P}rogram} & yes\\ 
     & \textit{Check} & \textit{none}, \textit{\textbf{U}ser}, \textit{\textbf{L}LM}, \textit{\textbf{P}rogram} & enforced \\ 
     \hline
     & \textit{Skills} & \textit{re\textbf{W}rite}, \textit{\textbf{C}reate}, \textit{con\textbf{V}erse}, \textit{\textbf{I}nform}, \textit{\textbf{R}eason}, \textit{\textbf{P}lan} & no \\
     \hline
     \textit{Output} & \textit{Format} & \textit{\textbf{F}reeText}, \textit{\textbf{I}tem}, \textit{\textbf{C}ode}, \textit{\textbf{S}tructure} & no \\
      & \textit{Revision} & \textit{none}, \textit{\textbf{U}ser}, \textit{\textbf{L}LM}, \textit{\textbf{P}rogram} & enforced \\ 
     & \textit{Consumer} & \textit{\textbf{U}ser}, \textit{\textbf{L}LM}, \textit{\textbf{P}rogram}, \textit{\textbf{E}ngine} & enforced \\
\hline  
\end{tabularx}
\end{table*}

\subsection{Overview and demonstration}

The taxonomy identifies 13 dimensions for LLM components, grouped into five metadimensions as shown in \cref{tab:dimover}. It comprises both dimensions with genuinely mutually exclusive characteristics and those with non-exclusive characteristics. 
For dimensions related to the technical integration of LLMs within applications, mutual exclusiveness is enforced.
Given the open nature of software architecture, the integration of LLMs allows
for significant diversity. In practice, LLM components may show multiple
characteristics within these dimensions. Nonetheless, the taxonomy requires
categorizing each component with a predominant characteristic, enforcing a
necessary level of abstraction to effectively organize and structure the domain. 

\begin{figure*}[t!]
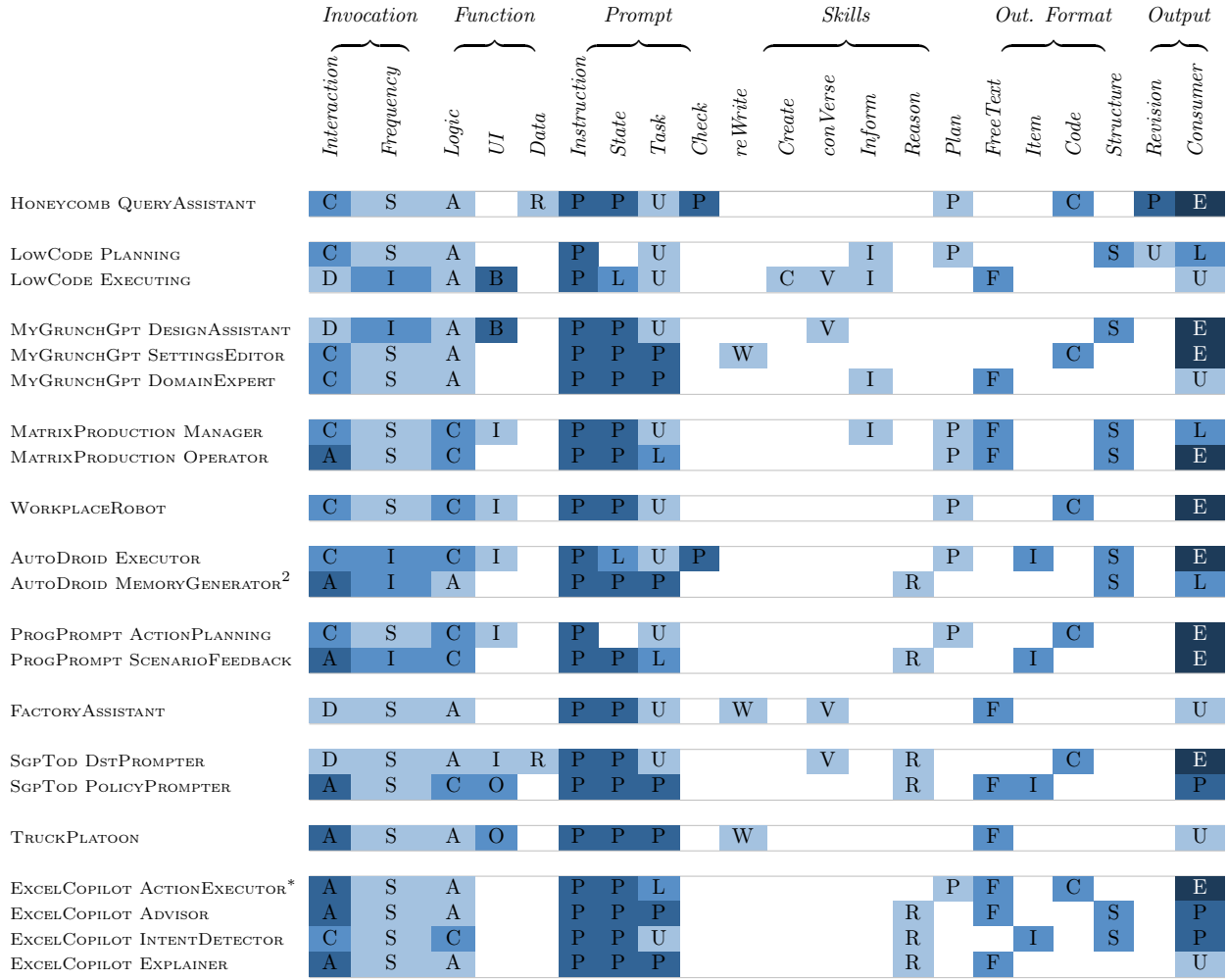

\resizebox{\textwidth}{!}{%
\begin{tabular}{@{}lccccccccccccccccccccc}
&\multicolumn{2}{c}{\textit{Invocation}} & \multicolumn{3}{c}{\textit{Function}} & \multicolumn{4}{c}{\textit{Prompt}} & \multicolumn{6}{c}{\textit{Skills}}& \multicolumn{4}{c}{\textit{Out. Format}}&\multicolumn{2}{c}{\textit{Output}} \\

&  \multicolumn{2}{c}{\downbracefill}   & \multicolumn{3}{c}{\downbracefill} & \multicolumn{4}{c}{\downbracefill} & \multicolumn{6}{c}{\downbracefill}& \multicolumn{4}{c}{\downbracefill}&\multicolumn{2}{c}{\downbracefill} \\

 &  \rotatebox{90}{\textit{Interaction}} &\rotatebox{90}{\textit{Frequency}} & \rotatebox{90}{\textit{Logic}} & \rotatebox{90}{\textit{UI }} & \rotatebox{90}{\textit{Data }} & \rotatebox{90}{\textit{Instruction}} & \multicolumn{1}{l}{\rotatebox{90}{\textit{State}}} & \multicolumn{1}{l}{\rotatebox{90}{\textit{Task}}} & \multicolumn{1}{l}{\rotatebox{90}{\textit{Check}}} & \multicolumn{1}{l}{\rotatebox{90}{\textit{reWrite}}} & \multicolumn{1}{l}{\rotatebox{90}{\textit{Create}}} & \multicolumn{1}{l}{\rotatebox{90}{\textit{conVerse}}} & \multicolumn{1}{l}{\rotatebox{90}{\textit{Inform}}} & \multicolumn{1}{l}{\rotatebox{90}{\textit{Reason}}} & \multicolumn{1}{l}{\rotatebox{90}{\textit{Plan}}}  & \multicolumn{1}{l}{\rotatebox{90}{\textit{FreeText}}} & \multicolumn{1}{l}{\rotatebox{90}{\textit{Item}}} & \multicolumn{1}{l}{\rotatebox{90}{\textit{Code}}}& \multicolumn{1}{l}{\rotatebox{90}{\textit{Structure}}} & \multicolumn{1}{l}{\rotatebox{90}{\textit{Revision}}} & \multicolumn{1}{l}{\rotatebox{90}{\textit{Consumer}}}   \\

&  &&&    &  &    &&&&    && &&  &  &   &&&  &  &    \\
\cline{2-22} \cline{2-22}
  \textsc{\footnotesize Honeycomb QueryAssistant}& \cellcolor{C}C& {\cellcolor{sing}S}  & {\cellcolor{calc}A}    &&\cellcolor{R}R  & \cellcolor{P}P  & \cellcolor{P}P   & \cellcolor{U}U  & \cellcolor{P}P   &      &        &       &        &        & \cellcolor{D}P &    &    & \cellcolor{form}C    && \cellcolor{P}P  & \cellcolor{E}{\color[HTML]{FFFFFF} E} \\ 
\cline{2-22}
&  &&&    &  &    &&&&    && &&  &  &   &&&  &  &   \\
\cline{2-22}
    \textsc{\footnotesize LowCode Planning}    & \cellcolor{C}C& 
  {\cellcolor{sing}S}  & {\cellcolor{calc}A}    && & \cellcolor{P}P  &    & \cellcolor{U}U  &    &      &        &       & \cellcolor{D}I  &        & \cellcolor{D}P &    &    &    & \cellcolor{form}S  & \cellcolor{U}U  & \cellcolor{L}L      \\
    \textsc{\footnotesize LowCode Executing}    & \cellcolor{D}D& \cellcolor{Iter}I    & {\cellcolor{calc}A}    & \cellcolor{B}B  &  & \cellcolor{P}P  & \cellcolor{L}L   & \cellcolor{U}U  &    &      & \cellcolor{D}C  & \cellcolor{D}V    & \cellcolor{D}I  &        &       & \cellcolor{form}F &    &    &&  & \cellcolor{U}U      \\
\cline{2-22}
&  &&&    &  &    &&&&    && &&  &  &   &&&  &  &   \\
\cline{2-22}
    \textsc{\footnotesize MyGrunchGpt DesignAssistant}  & \cellcolor{D}D& \cellcolor{Iter}I    & {\cellcolor{calc}A}    & \cellcolor{B}B  &  & \cellcolor{P}P  & \cellcolor{P}P   & \cellcolor{U}U  &    &      &        & \cellcolor{D}V    &        &        &       &    &    &    & \cellcolor{form}S  &      & \cellcolor{E}{\color[HTML]{FFFFFF} E} \\
    \textsc{\footnotesize MyGrunchGpt SettingsEditor}   & \cellcolor{C}C& {\cellcolor{sing}S}  & {\cellcolor{calc}A}    && & \cellcolor{P}P  & \cellcolor{P}P   & \cellcolor{P}P  &    & \cellcolor{D}W   &        &       &        &        &       &    &    & \cellcolor{form}C    &&  & \cellcolor{E}{\color[HTML]{FFFFFF} E} \\
    \textsc{\footnotesize MyGrunchGpt DomainExpert}& \cellcolor{C}C& {\cellcolor{sing}S}  & {\cellcolor{calc}A}    && & \cellcolor{P}P  & \cellcolor{P}P   & \cellcolor{P}P  &    &      &        &       & \cellcolor{D}I  &        &       & \cellcolor{form}F &    &    &&  & \cellcolor{U}U      \\
\cline{2-22}
&  &&&    &  &    &&&&    && &&  &  &   &&&  &  &   \\
\cline{2-22}
   \textsc{\footnotesize  MatrixProduction Manager}& \cellcolor{C}C& {\cellcolor{sing}S}  & \cellcolor{Con}C   & \cellcolor{I}I  &  & \cellcolor{P}P  & \cellcolor{P}P   & \cellcolor{U}U  &    &      &        &       & \cellcolor{D}I &        & \cellcolor{D}P & \cellcolor{form}F &    &    & \cellcolor{form}S  &  & \cellcolor{L}L      \\
    \textsc{\footnotesize MatrixProduction Operator}   & \cellcolor{A}A& {\cellcolor{sing}S}  & \cellcolor{Con}C   && & \cellcolor{P}P  & \cellcolor{P}P   & \cellcolor{L}L  &    &      &        &       &        &        & \cellcolor{D}P & \cellcolor{form}F &    &    & \cellcolor{form}S  &      & \cellcolor{E}{\color[HTML]{FFFFFF} E} \\
\cline{2-22}
&  &&&    &  &    &&&&    && &&  &  &   &&&  &  &   \\
\cline{2-22}
    \textsc{\footnotesize WorkplaceRobot}  & \cellcolor{C}C& {\cellcolor{sing}S}  & \cellcolor{Con}C   & \cellcolor{I}I  &  & \cellcolor{P}P  & \cellcolor{P}P   & \cellcolor{U}U  &    &      &        &       &        &        & \cellcolor{D}P &    &    & \cellcolor{form}C    &&      & \cellcolor{E}{\color[HTML]{FFFFFF} E} \\
\cline{2-22}
&  &&&    &  &    &&&&    && &&  &  &   &&&  &  &   \\
\cline{2-22}
    \textsc{\footnotesize AutoDroid Executor}   & \cellcolor{C}C& \cellcolor{Iter}I    & \cellcolor{Con}C   & \cellcolor{I}I  &  & \cellcolor{P}P  & \cellcolor{L}L   & \cellcolor{U}U  & \cellcolor{P}P   &      &        &       &        &        & \cellcolor{D}P &    &  \cellcolor{form}I  &   & \cellcolor{form}S  &      & \cellcolor{E}{\color[HTML]{FFFFFF} E}  \\
    \textsc{\footnotesize AutoDroid MemoryGenerator}$^2$  & \cellcolor{A}A& \cellcolor{Iter}I    & {\cellcolor{calc}A}    && & \cellcolor{P}P  & \cellcolor{P}P   & \cellcolor{P}P  &    &      &        &       &        & \cellcolor{D}R  &       &    &    &    & \cellcolor{form}S  &  & \cellcolor{L}L      \\
\cline{2-22}
&  &&&    &  &    &&&&    && &&  &  &   &&&  &  &   \\
\cline{2-22}
   \textsc{\footnotesize  ProgPrompt ActionPlanning}    & \cellcolor{C}C& {\cellcolor{sing}S}  & \cellcolor{Con}C   & \cellcolor{I}I  &  & \cellcolor{P}P  &    & \cellcolor{U}U  &    &      &        &       &        &        & \cellcolor{D}P &    &    & \cellcolor{form}C    &&  & \cellcolor{E}{\color[HTML]{FFFFFF} E} \\
   \textsc{\footnotesize  ProgPrompt ScenarioFeedback}  & \cellcolor{A}A& \cellcolor{Iter}I    & \cellcolor{Con}C   && & \cellcolor{P}P  & \cellcolor{P}P   & \cellcolor{L}L  &    &      &        &       &        & \cellcolor{D}R  &       &    & \cellcolor{form}I    &    &&  & \cellcolor{E}{\color[HTML]{FFFFFF} E} \\
\cline{2-22}
&  &&&    &  &    &&&&    && &&  &  &   &&&  &  &   \\
\cline{2-22}
    \textsc{\footnotesize FactoryAssistant}& \cellcolor{D}D& {\cellcolor{sing}S}  & {\cellcolor{calc}A}    && & \cellcolor{P}P  & \cellcolor{P}P   & \cellcolor{U}U  &    & \cellcolor{D}W   &        & \cellcolor{D}V    &        &        &       & \cellcolor{form}F &    &    &&  & \cellcolor{U}U      \\
\cline{2-22}
&  &&&    &  &    &&&&    && &&  &  &   &&&  &  &   \\
\cline{2-22}
   \textsc{\footnotesize  SgpTod DstPrompter}   & \cellcolor{D}D& {\cellcolor{sing}S}  & {\cellcolor{calc}A}    & \cellcolor{I}I  & \cellcolor{R}R  & \cellcolor{P}P  & \cellcolor{P}P   & \cellcolor{U}U  &    &      &        & \cellcolor{D}V    &        & \cellcolor{D}R  &       &    &    & \cellcolor{form}C    &&  & \cellcolor{E}{\color[HTML]{FFFFFF} E} \\
    \textsc{\footnotesize SgpTod PolicyPrompter}   & \cellcolor{A}A& {\cellcolor{sing}S}  & \cellcolor{Con}C   & \cellcolor{O}O  &  & \cellcolor{P}P  & \cellcolor{P}P   & \cellcolor{P}P  &    &      &        &       &        & \cellcolor{D}R  &       & \cellcolor{form}F & \cellcolor{form}I    &    &&  & \cellcolor{P}P      \\
\cline{2-22}
&  &&&    &  &    &&&&    && &&  &  &   &&&  &  &   \\
\cline{2-22}
   \textsc{\footnotesize TruckPlatoon}    & \cellcolor{A}A& {\cellcolor{sing}S}  & {\cellcolor{calc}A}    & \cellcolor{O}O  &  & \cellcolor{P}P  & \cellcolor{P}P   & \cellcolor{P}P  &    & \cellcolor{D}W   &        &       &        &        &       & \cellcolor{form}F &    &    &&  & \cellcolor{U}U      \\
\cline{2-22}
&  &&&    &  &    &&&&    && &&  &  &   &&&  &  &   \\
\cline{2-22}
    \textsc{\footnotesize ExcelCopilot ActionExecutor}$^{\ast}$ & \cellcolor{A}A& {\cellcolor{sing}S}  & {\cellcolor{calc}A}    && & \cellcolor{P}P  & \cellcolor{P}P   & \cellcolor{L}L  &    &      &        &       &        &        & \cellcolor{D}P & \cellcolor{form}F &    & \cellcolor{form}C    &&  & \cellcolor{E}{\color[HTML]{FFFFFF} E} \\
    \textsc{\footnotesize ExcelCopilot Advisor}    & \cellcolor{A}A& {\cellcolor{sing}S}  & {\cellcolor{calc}A}    && & \cellcolor{P}P  & \cellcolor{P}P   & \cellcolor{P}P  &    &      &        &       &        & \cellcolor{D}R  &       & \cellcolor{form}F &    &    & \cellcolor{form}S  &  & \cellcolor{P}P      \\
    \textsc{\footnotesize ExcelCopilot  IntentDetector} & \cellcolor{C}C& {\cellcolor{sing}S}  & \cellcolor{Con}C   && & \cellcolor{P}P  & \cellcolor{P}P   & \cellcolor{U}U  &    &      &        &       &        & \cellcolor{D}R  &       &    &   \cellcolor{form}I  &    & \cellcolor{form}S  &  & \cellcolor{P}P      \\
   \textsc{\footnotesize ExcelCopilot Explainer}  & \cellcolor{A}A& {\cellcolor{sing}S}  & {\cellcolor{calc}A}    && & \cellcolor{P}P  & \cellcolor{P}P   & \cellcolor{P}P  &    &      &        &       &        & \cellcolor{D}R  &       & \cellcolor{form}F &    &    &&  & \cellcolor{U}U       \\
\cline{2-22}
\end{tabular}}
\caption{Categorized example instances. See \cref{tab:dimover} for a legend. $\ast, 2$: multiple LLM components.}\label{fig:taxsum}
\end{figure*}

We applied the taxonomy to categorize each of the example instances described
in \cref{sec:cases}. The results are depicted in \cref{fig:taxsum}. The dimensions and their characteristics are detailed and illustrated with examples in \cref{sec:characteristics}.

The taxonomy visualizes an LLM component by a feature vector
comprising binary as well as multivalued features. Non-mutually exclusive
dimensions are represented by a set of binary features. The remaining
dimensions are encoded as $n$-valued features where $n$ denotes the number of characteristics. For compactness, we use one-letter codes of the characteristics
as feature values in the visualizations. In \cref{tab:dimover}, these codes are
printed in upper case in the respective characteristic's name. 

A feature vector representing an LLM component is visualized in one line. For dimensions with non-mutually exclusive characteristics, all possible codes are listed, with the applicable ones marked. The remaining dimensions are represented by the code of the applicable characteristic, with the characteristic \textit{none} shown as an empty cell. We shade feature values  with different tones to support visual perception. LLM components within the same application are grouped
together, visualizing an LLM-integrating application in a tabular format.

\subsection{Dimensions and characteristics}\label{sec:characteristics}

\subsubsection{\textit{Invocation} dimensions}

Two \textit{Invocation} dimensions address the way the LLM is invoked within
the application. 

\mydimpar{\textit{Interaction}} describes how the user interacts with
the LLM with three characteristics:

\textit{App}: Users never converse with the LLM directly in natural language,
rather the application invokes the LLM automatically. \Myeg users do not
interact directly with \textsc{ExcelCopilot ActionExecutor} or with
\textsc{MatrixProduction Operator}.

\textit{Command}: Users input single natural language commands. \Myeg users interact with \textsc{AutoDroid TaskExecutor} through single natural language commands.

\textit{Dialog}: Users engage in multi-turn dialogues with the LLM component to achieve a use goal. \Myeg users repeatedly prompt \textsc{LowCode Executing} or \textsc{MyCrunchGpt DesignAssistant} in multi-turn dialogues to obtain an essay or an airfoil design, respectively.

\mydimpar{\textit{Frequency}} addresses how often the application
invokes a specific LLM component to fulfill a goal:

\textit{Single}: A single invocation of an LLM component is sufficient to produce the result. \Myeg in \textsc{MyCrunchGpt}, the application internally invokes distinct LLM components once for each user input by injecting varying prompt instructions.

\textit{Iterative}: The LLM component is invoked repeatedly to produce the result. \Myeg \textsc{AutoDroid TaskExecutor} is invoked multiple times to fulfill a command with an updated environment description in the \textit{State} prompt; \textsc{LowCode Executing} is repeatedly prompted by the user to achieve the use goal while the application updates the dialogue history.

\subsubsection{\textit{Function} dimensions}
 The \textit{Function} dimensions are derived from the classical three-tier software architecture model which segregates an application into three distinct layers: presentation, logic and data \cite{fowlerPatternsEnterpriseApplication2002}. The presentation layer implements the UI. On the input side, it allows users to enter data and commands that control the application. On the output side, it presents information and provides feedback on the execution of commands. The logic layer holds the code that directly realizes the core objectives and processes of an application such as processing data, performing calculations, and making decisions. The data layer of an application manages the reading and writing of data from and to persistent data storage.
Due to its versatility, an LLM component can simultaneously implement functionality for all three layers. The taxonomy addresses this with three \textit{Function} dimensions.

\mydimpar{\textit{UI}} indicates whether an LLM component contributes significantly to the user interface of an application, avoiding the need to implement graphical UI controls or display elements:

\textit{none}: No UI functionality is realized by the LLM. \Myeg in \textsc{ExcelCopilot}, the LLM does not replace any UI elements.

\textit{Input}: Input UI is (partially) implemented by the LLM. \Myeg	in \textsc{MatrixProduction Manager}, users input their order in natural language, obviating a product configuration GUI.

\textit{Output}: Output UI is (partially) implemented by the LLM. \Myeg in \textsc{TruckPlatoon}, the output generated by the LLM component can replace a data cockpit with gauges and other visuals displaying numerical data.

\textit{Both}:	 Input and output UI are (partially) implemented by the LLM. \Myeg in \textsc{MyCrunchGpt}, the \textsc{DesignAssistant} provides a convenient conversational interface for parameterization of APIs and tools and feedback on missing values, which otherwise might require a complex GUI.

\mydimpar{\textit{Logic}} indicates whether the LLM component determines the control flow of the application. It discerns two characteristics:

\textit{cAlculate}: The output does not significantly impact the control flow of the application, \myie the output is processed like data. \Myeg \textsc{MyCrunchGpt SettingsEditor} modifies a JSON file, replacing a programmed function; \textsc{MyCrunchGpt DesignAssistant} asks the user for 
parameters, but the sequence of calling APIs and tools follows a predefined workflow; the workflow computed by \textsc{LowCode Planning} is displayed without influencing the application's control flow.

\textit{Control}: 	The output of the LLM is used for controlling the application. \Myeg
the plans generated by \textsc{MatrixProduction Manager} serve to schedule and activate production modules; the actions proposed by \textsc{AutoDroid TaskExecutor} are actually executed and determine how the control flow of the app proceeds.

Since an LLM invocation always computes a result, \textit{cAlculate} is interpreted as ``calculate only'', making \textit{cAlculate} and \textit{Control} mutually exclusive.

\mydimpar{\textit{Data}} addresses whether the LLM contributes to reading or writing persistent data:

\textit{none}: The LLM does not contribute to reading or writing persistent data. This characteristic applies to most sample instances. 

\textit{Read}: 	The LLM is applied for reading from persistent data store. \Myeg \textsc{Sgp\-Tod DstPrompter} generates SQL queries which the application executes;
\textsc{Honeycomb QueryAssistant} devises analytical database queries.

\textit{Write} and \textit{Both}: 	 
No LLM component among the samples generates
database queries for creating or updating persistent data.

\subsubsection{Prompt-related dimensions}
Integrating an LLM into an application poses specific requirements for prompts, such as the need for prompts to reliably elicit output in the requested form \cite{xiaFOFOBenchmarkEvaluate2024}. While a broad range of prompt patterns have been identified and investigated \cite{whitePromptPatternCatalog2023a}, there is still a lack of research on successful prompt patterns specifically for LLM-integrated applications, on which this taxonomy could build. Developing prompt taxonomies is a challenging research endeavor in itself \cite{santuTELeRGeneralTaxonomy2023} and is beyond the scope of this research. Therefore, the taxonomy does not define a dimension with specific prompt patterns as characteristics, but rather focuses on how the application generates the prompt for an LLM component from a technical perspective. 

Prompts generally consist of several parts with distinct purposes, generated by
different mechanisms. Although many authors explore the concepts,
a common terminology has yet to be established. This is illustrated in
\cref{tab:promptparts}, showing terms from an ad-hoc selection of recent papers addressing prompt generation in applications. In the table, italics
indicate that the authors refrain from introducing an abstract term and instead
use a domain-specific description. The term ``examples'' indicates a
\textit{one-shot} or \textit{few-shot} prompt pattern. The terms that are
adopted for the taxonomy are underlined. 

\begin{table*}[tbh]
 \caption{Terms used for prompt parts. Expressions specific to a domain  are printed in italics, ``examples'' indicates a
\textit{one-shot} or \textit{few-shot} prompt pattern. Terms adopted for the taxonomy are underlined.}
 \label{tab:promptparts}
\medskip

\noindent
\begin{tabularx}{\textwidth}{lp{6cm}Xl}
\hline
Source & Instruction & State & Task\\
\hline
\cite{zhaoSurveyLargeLanguage2023} & task description + examples & ~ & test
instance\\ 
\cite{liuPromptInjectionAttacks2023} & \underline{instruction} prompt & ~ & data
prompt\\
\cite{liuPromptInjectionAttack2023} & predefined prompt & ~ & user
prompt\\
\cite{pedroPromptInjectionsSQL2023} & \raggedright{prompt template + examples }& \emph{DB schema} &
user input question\\
\cite{pedroPromptInjectionsSQL2023} & examples & ~ & \textit{SQL query result}\\
\cite{meesGroundingLanguageVisual2023} & prompt context, \myie examples & \raggedright{\emph{environment \underline{state}, scene description}} & input \underline{task} commands\\
\cite{caiLowcodeLLMVisual2023} & education prompt & dialogue history & user input \underline{task} prompt\\
\cite{caiLowcodeLLMVisual2023} & education prompt & \raggedright{dialogue history +
\emph{provided workflow}}& \emph{(circumscribed)}\\
\cite{XiaAutonomousSystemFlexible2023} & role and goal + \underline{instruction} + examples &
context & current \underline{task}\\
\cite{kernanfreireHarnessingLargeLanguage2023} & \raggedright{predefined system \underline{instruction} +
\emph{domain-specific information}} & \raggedright{\emph{query results from 
knowledge graph}} & the user's request\\
\hline
\end{tabularx}
\end{table*}

The taxonomy distinguishes three prompt parts referred to as \textit{Prompt
Instruction}, \textit{Prompt State}, and \textit{Prompt Task}. These parts can occur in any order, potentially interleaved, and some parts may be absent. 
\begin{itemize}
\item \textit{Instruction} is the part of a prompt that outlines how to solve the task. Defined during LLM component development, it remains static throughout an application's lifespan. 
\item \textit{State} is the situation-dependent part of the prompt that is created dynamically every time the LLM is invoked. The taxonomy opts for the term \textit{State} instead of ``context'' in order to avoid confusion with the ``LLM context'' as explained in \cref{sec:background}. The \textit{State} may include the current dialogue history, an extract of a knowledge base needed specifically for the current LLM invocation, or a state or scene description, etc. 
\item \textit{Task} is the part of the prompt conveying the task to solve in a specific invocation. 
\end{itemize}

\mydimpar{\textit{Prompt Instruction}, \textit{State} and \textit{Task}} describe the origins 
of the prompt parts by uniform characteristics:

\textit{none}: 	The prompt part is not present. \Myeg \textsc{ProgPrompt ActionPlanning} has no \textit{State} prompt, nor does \textsc{LowCode Planning} (except the dialogue history when planning a subprocess). \textit{Instruction} and \textit{Task} prompt parts are present in all sample instances. 

\textit{User}: The user phrases the prompt part. \Myeg the \textit{Task} for \textsc{ExcelCopilot IntentDetector} or for \textsc{LowCode Planning} is phrased by the user. There are no sample instances where the user provides the \textit{Instruction} or \textit{State} prompt parts. 
 
\textit{LLM}: The prompt part is generated by an LLM. \Myeg \textsc{LowCode Planning} generates the \textit{State} for \textsc{LowCode Executing} and \textsc{ExcelCopilot IntentDetector} generates the \textit{Task} for \textsc{ExcelCopilot ActionExecutor}s.

\textit{Program}: Application code generates the prompt part. \Myeg \textsc{AutoDroid} programmatically generates 
 the \textit{State} and the \textit{Task} parts for its \textsc{MemoryGenerator}s in the knowledge base building phase.
 
The \textit{Prompt Instruction} dimension is always generated by \textit{Program}. While a user and possibly an LLM have defined this prompt part during application development, this falls outside the scope of this taxonomy. Therefore, the \textit{Prompt Instruction} dimension is not discriminating and categorizes all cases as \textit{Program}. It is retained in the taxonomy for completeness and better understandability.

\mydimpar{\textit{Prompt Check}} describes whether the application employs a
review mechanism to control and modify the prompt before invoking the LLM.
The same characteristics as for the prompt parts are applicable: 

\textit{none}: 	The prompt is used without check. 

\textit{User}: The user checks and revises the prompt. 

\textit{LLM}: 	Another LLM component checks or revises the prompt.  

\textit{Program}:	 The application comprises code to check or revise the prompt. \Myeg \textsc{AutoDroid} removes personal data, such as names, to ensure privacy before invoking the \textsc{TaskExecutor}; \textsc{Honeycomb QueryAssistant} incorporates a coded mechanism against prompt injection attacks.

Most example instances omit prompt checks. There are no examples where a \textit{Check} is performed by a \textit{User} or an \textit{LLM}.

\subsubsection{\textit{Skills} dimensions}

The \textit{Skills} dimension captures the types of LLM capabilities that an application utilizes. It is designed as a dimension with six non-mutually exclusive characteristics. 

\begin{samepage}
\mydimpar{\textit{Skills}} is decomposed into six specific capabilities:

\textit{reWrite}: The LLM edits or transforms data or text, such as rephrasing, summarizing, reformatting, correcting, or replacing values. \Myeg \textsc{MyCrunchGpt SettingsEditor} replaces values in JSON files; \textsc{TruckPlatoon} converts measurements into textual explanations.
\end{samepage}

\textit{Create}: The LLM generates novel output. \Myeg \textsc{LowCode Executing} generates substantial bodies of text for tasks like essay writing.

\textit{conVerse}: The application relies on the LLM’s capability to engage in purposeful dialogues with humans. \Myeg \textsc{MyCrunchGpt DesignAssistant} asks users for missing parameters; \textsc{Sgp\-Tod PolicyPrompter} decides how to react to user inputs and formulates chatbot responses.

\textit{Inform}: The application depends on knowledge that the LLM has acquired during its training, unlike applications that provide all necessary information within the prompt. \Myeg \textsc{MyCrunchGpt DomainExpert} provides expert knowledge on airfoil designs; \textsc{MatrixProduction} relies on built-in knowledge of production processes, such as ``a hole is produced by drilling''; \textsc{LowCode Executing} uses its learned knowledge for tasks like essay writing.

\textit{Reason}: The LLM draws conclusions or makes logical inferences. \Myeg \textsc{FormulaExplainer} in \textsc{ExcelCopilot} explains the effects of Excel functions in formulas; \textsc{AutoDroid MemoryGenerator}s explain the effects of GUI elements in Android apps. 

\textit{Plan}: The LLM designs a detailed method or course of action to achieve a specific goal. \Myeg \textsc{AutoDroid TaskExecutor} and \textsc{Workplace\-Robot TaskPlanning} devise action plans to achieve goals.

The \textit{Plan} and \textit{Reason} characteristics are interrelated, as planning also requires reasoning. The intended handling of these characteristics is to categorize an LLM component as \textit{Plan} only and understand \textit{Plan} as implicitly subsuming \textit{Reason}.

The effectiveness of LLMs as components of software applications relies on their commonsense knowledge and their ability to correctly interpret and handle a broad variety of text inputs, including instructions, examples, and code. It is reasonable to assume that a fundamental capability, which might be termed \textit{Unterstand}, is leveraged by every LLM component. As it is not distinctive, the taxonomy does not list it explicitly in the \textit{Skills} dimension. 

Applying this taxonomy dimension requires users to determine which skills are most relevant and worth highlighting in an LLM component. Given the versatility of LLMs, reducing the focus to few predominant skills is necessary to make categorizations  distinctive and expressive. 

\subsubsection{Output-related dimensions}\label{output-related-dimensions}
\mydimpar{\textit{Output Format}} characterizes the format of the LLM’s output. As an output may consist of several parts in diverse formats, this dimension is designed as non-mutually exclusive, same as the \textit{Skills} dimension.
It distinguishes four characteristics that are distinctive and well discernible:
 
\textit{FreeText}: unstructured natural language text output. \Myeg \textsc{TruckPlatoon} and \textsc{MyCrunchGpt DomainExpert} generate text output in natural language; \textsc{MatrixProduction Manager} and \textsc{MatrixProduction Operator} produce \textit{FreeText} explanations complementing output in custom formats to be parsed by the application.

\textit{Item}: a single text item from a predefined set of items, such as a class in a classification task. \Myeg \textsc{ProgPrompt ScenarioFeedback} outputs either \texttt{True} or \texttt{False}.

\textit{Code}: source code or other highly formalized output that the LLM has learned during its training, such as a programming language, XML, or JSON. \Myeg \textsc{AutoDroid TaskExecutor} produces code to steer an Android app; \textsc{MyCrunchGpt SettingsEditor} outputs JSON.

\textit{Structure}: structured, formalized output adhering to a custom format. \Myeg \textsc{LowCode Planning} outputs text in a format that can be displayed as a flow chart; \textsc{MatrixProduction Manager} and \textsc{Operator} produce output in custom formats combined with \textit{FreeText} explanations.

\mydimpar{\textit{Output Revision}} indicates whether the application checks or revises the LLM-generated output before utilization. These characteristics and their interpretations mirror those in the \textit{Prompt Check} dimension:

\textit{none}: There is no revision of the LLM output. 

\textit{User}: The user revises the LLM output. \Myeg the user improves the plan generated by \textsc{LowCode Planning}.

\textit{LLM}: A further LLM component checks or revises the output of the LLM component under consideration.

\textit{Program}: Programmed code checks or revises the LLM output. \Myeg \textsc{Honeycomb QueryAssistant} corrects the query produced by the LLM before executing it \cite{carterAllHardStuff2023}.

There are no instances in the sample set where another LLM revises or checks the output of the LLM.
Most sample applications do not check or revise the LLM's output, though several of them parse and transform it. 
The purpose of the \textit{Output Revision} dimension is to indicate whether the application includes control or correction mechanisms, rather than just parsing it.

\mydimpar{\textit{Output Consumer}} addresses the way of utilizing the LLM output:

\textit{User} signifies that the LLM output is presented to a human user. \Myeg the text output of \textsc{TruckPlatoon} is intended for humans, as well as the output of \textsc{MyCrunchGPT DomainExpert}. 

\textit{LLM} indicates that the output serves as a prompt part in a further LLM invocation. \Myeg the knowledge base entries generated by an \textsc{AutoDroid MemoryGenerator} become part of the prompt for \textsc{AutoDroid TaskExecutor}; the plan output by \textsc{LowCode Planning} serves as a part of the prompt for \textsc{LowCode Executing}. 

\textit{Program} describes instances where the LLM output is consumed and  processed further by a software component of the application. \Myeg the output of \textsc{MatrixProduction Manager} is handled by software systems (including a Manufacturing Execution System) which use it to compute prompts for other LLM components.

\textit{Engine} covers scenarios where the LLM output is intended for execution on a runtime engine. \Myeg the SQL query generated by \textsc{Sgp\-Tod DstPrompter} is processed by a SQL interpreter; a part of the output of \textsc{MatrixProduction Operator} is executed by automation modules.

Although applications may parse and transform the LLM output before use, the \textit{Output Consumer} dimension is meant to identify the ultimate consumer, such as an execution engine, rather than an intermediary parser or transformation code. When applications divide the LLM output into parts for different consumers, users applying the taxonomy need to determine which consumer is most relevant, since this dimension is designed to be mutually exclusive.

\subsection{Evaluation}\label{sec:eval-taxcomp}

\begin{figure*}[thb]
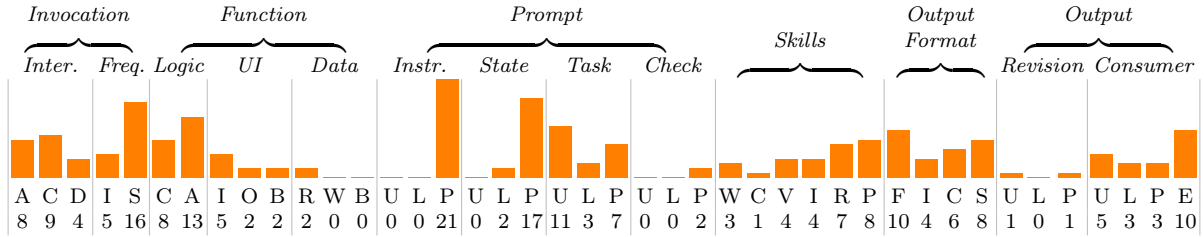

 \setlength{\tabcolsep}{1pt}
\footnotesize \center
\begin{tabular}{|ccc|cc|cc|ccc|ccc|ccc|ccc|ccc|ccc|cccccc|cccc|ccc|cccc|}

\multicolumn{5}{c}{\textit{Invocation}} &
\multicolumn{8}{c}{\textit{Function}} &

\multicolumn{12}{c}{\textit{Prompt }}&

\multicolumn{6}{c}{\textit{ }} &

\multicolumn{4}{c}{\textit{Output}} &
\multicolumn{7}{c}{\textit{Output}} 

\\
\multicolumn{5}{c}{\downbracefill} &
\multicolumn{8}{c}{\downbracefill} &
\multicolumn{12}{c}{\downbracefill}&
\multicolumn{6}{c}{\textit{Skills}} &
\multicolumn{4}{c}{\textit{Format}} &
\multicolumn{7}{c}{\downbracefill} 
\\

\multicolumn{3}{c}{\textit{Inter. }} &
\multicolumn{2}{c}{\textit{Freq. }} &

\multicolumn{2}{c}{\textit{Logic }} &
\multicolumn{3}{c}{\textit{UI }} &
\multicolumn{3}{c}{\textit{Data }} &

\multicolumn{3}{c}{\textit{Instr. }}&
\multicolumn{3}{c}{\textit{State}} &
\multicolumn{3}{c}{\textit{Task}} &
\multicolumn{3}{c}{\textit{Check }}&

\multicolumn{6}{c}{\downbracefill} &
\multicolumn{4}{c}{\downbracefill} &

\multicolumn{3}{c}{\textit{Revision}} &
\multicolumn{4}{c}{\textit{Consumer }} \\

\tikz \fill [orange] (0.1,0.1) rectangle (0.4,0.6); &
\tikz \fill [orange] (0.1,0.1) rectangle (0.4,0.66); &
\tikz \fill [orange] (0.1,0.1) rectangle (0.4,0.35); &

\tikz \fill [orange] (0.1,0.1) rectangle (0.4,0.41); &
\tikz \fill [orange] (0.1,0.1) rectangle (0.4,1.1); &

\tikz \fill [orange] (0.1,0.1) rectangle (0.4,0.6); &
\tikz \fill [orange] (0.1,0.1) rectangle (0.4,0.91); &

\tikz \fill [orange] (0.1,0.1) rectangle (0.4,0.41); &
\tikz \fill [orange] (0.1,0.1) rectangle (0.4,0.23); &
\tikz \fill [orange] (0.1,0.1) rectangle (0.4,0.23); &

\tikz \fill [orange] (0.1,0.1) rectangle (0.4,0.23); &
\tikz \fill [orange] (0.1,0.1) rectangle (0.4,0.1); &
\tikz \fill [orange] (0.1,0.1) rectangle (0.4,0.1); &

\tikz \fill [orange] (0.1,0.1) rectangle (0.4,0.1); &
\tikz \fill [orange] (0.1,0.1) rectangle (0.4,0.1); &
\tikz \fill [orange] (0.1,0.1) rectangle (0.4,1.41); &

\tikz \fill [orange] (0.1,0.1) rectangle (0.4,0.1); &
\tikz \fill [orange] (0.1,0.1) rectangle (0.4,0.23); &
\tikz \fill [orange] (0.1,0.1) rectangle (0.4,1.16); &

\tikz \fill [orange] (0.1,0.1) rectangle (0.4,0.79); &
\tikz \fill [orange] (0.1,0.1) rectangle (0.4,0.29); &
\tikz \fill [orange] (0.1,0.1) rectangle (0.4,0.54); &

\tikz \fill [orange] (0.1,0.1) rectangle (0.4,0.1); &
\tikz \fill [orange] (0.1,0.1) rectangle (0.4,0.1); &
\tikz \fill [orange] (0.1,0.1) rectangle (0.4,0.23); &

\tikz \fill [orange] (0.1,0.1) rectangle (0.4,0.29); &	
\tikz \fill [orange] (0.1,0.1) rectangle (0.4,0.16); &	
\tikz \fill [orange] (0.1,0.1) rectangle (0.4,0.35); &	
\tikz \fill [orange] (0.1,0.1) rectangle (0.4,0.35); &	
\tikz \fill [orange] (0.1,0.1) rectangle (0.4,0.54); &	
\tikz \fill [orange] (0.1,0.1) rectangle (0.4,0.6); &

\tikz \fill [orange] (0.1,0.1) rectangle (0.4,0.73); &	
\tikz \fill [orange] (0.1,0.1) rectangle (0.4,0.35); &
\tikz \fill [orange] (0.1,0.1) rectangle (0.4,0.48); &	
\tikz \fill [orange] (0.1,0.1) rectangle (0.4,0.6); &

\tikz \fill [orange] (0.1,0.1) rectangle (0.4,0.16); &
\tikz \fill [orange] (0.1,0.1) rectangle (0.4,0.1); &
\tikz \fill [orange] (0.1,0.1) rectangle (0.4,0.16); &

\tikz \fill [orange] (0.1,0.1) rectangle (0.4,0.41); &
 \tikz \fill [orange] (0.1,0.1) rectangle (0.4,0.29); &
 \tikz \fill [orange] (0.1,0.1) rectangle (0.4,0.29); &
\tikz \fill [orange] (0.1,0.1) rectangle (0.4,0.73); 

\\
\footnotesize  A& \footnotesize C& \footnotesize  D & \footnotesize  I    & \footnotesize  S & \footnotesize  C    & \footnotesize  A & \footnotesize  I& \footnotesize  O & \footnotesize B & \footnotesize  R& \footnotesize  W & \footnotesize B & \footnotesize  U & \footnotesize L& \footnotesize  P & \footnotesize  U & \footnotesize L& \footnotesize  P & \footnotesize  U & \footnotesize L & \footnotesize P & \footnotesize  U & \footnotesize L & \footnotesize P & \footnotesize  W & \footnotesize  C & \footnotesize  V & \footnotesize  I & \footnotesize  R & \footnotesize  P & \footnotesize  F  & \footnotesize  I & \footnotesize  C & \footnotesize  S & \footnotesize  U& \footnotesize  L& \footnotesize  P & \footnotesize  U& \footnotesize L& \footnotesize P& \footnotesize E 
 \\
8 & 9 & 4 & 
 5 & 16 & 
 8 & 13 & 
 5 & 2 & 2 & 
 2 & 0 & 0 & 
 0 & 0 & 21 & 
 0 & 2 & 17 & 
 11 & 3 & 7 & 
 0 & 0 & 2 & 
 3 & 
 1 & 
 4 & 
 4 & 
 7 & 
 8 & 
 10 & 
 4 & 
 6 & 
 8 & 
 1 & 0 & 1 & 
 5 & 3 & 3 & 10 
\end{tabular}%
\normalsize
\caption{Occurrences of characteristics in the sample set of LLM-integrated applications.}\label{tab:taxcounts} 
\end{figure*}

\Cref{tab:taxcounts} displays the number of occurrences of characteristics within the example instances. It must be noted, however, that these do not reflect actual frequencies, as similar LLM components within the same application are aggregated together, indicated by symbols $*$ and $2$ in \cref{fig:taxsum}. Furthermore, \textsc{ExcelCopilot} likely includes occurrences of \textit{Prompt Check} and \textit{Output Revision} which are not counted due to insufficient system documentation.

We evaluate the taxonomy against commonly accepted quality criteria:
comprehensiveness, robustness, conciseness, mutual exclusiveness,
 explanatory power, and extensibility
\cite{unterkalmsteinerCompendiumEvaluationTaxonomy2023,nickersonMethodTaxonomyDevelopment2013}.
The taxonomy encompasses all example instances including those that were not
considered during its development. This demonstrates \textbf{comprehensiveness}.
As \cref{fig:taxsum} shows, all example instances have unique categorizations, supporting the
taxonomy's \textbf{robustness}. This not only indicates that the dimensions and characteristics are distinctive for the domain,
 but also highlights the wide variety possible in this field.
\textbf{Conciseness} demands that the taxonomy uses the minimum number of
dimensions and characteristics. The taxonomy gains conciseness by identifying
relatively few and abstract characteristics within each dimension. However, it
does not adhere to the related subcriterion that each characteristic must be
present in at least one investigated instance
\cite{szopinskiCriteriaPreludeGuiding2020}. Unoccupied characteristics are
retained for dimensions whose characteristics were derived conceptually,
specifically, for the \textit{Prompt} dimensions, the \textit{Output
Revision} dimension, and the \textit{Data Function} dimension, enhancing the
taxonomy's ability to illustrate design options and inspire novel uses for LLM
integrations in applications. Some dimensions are constructed in parallel, sharing common sets of characteristics. While this affects conciseness, it makes the taxonomy easier to understand and apply.
As is often seen in taxonomy development
\cite{szopinskiCriteriaPreludeGuiding2020}, we deliberately waived the requirement for \textbf{mutual
exclusiveness} for some dimensions, specifically the \textit{Output Format} and
\textit{Skills} dimensions. In the context of this taxonomy, these can equivalently be understood as a set of of six and four binary dimensions respectively, each divided into characteristics ``yes'' and ``no''. However, framing them as a single dimension with non-mutually exclusive characteristics seems more intuitive.

Metadimensions structure the taxonomy, and most of the characteristics are illustrated through examples. These measures are recognized for enhancing the \textbf{explanatory power} of a taxonomy 
\cite{unterkalmsteinerCompendiumEvaluationTaxonomy2023}.
The taxonomy's flat structure allows for the easy addition of dimensions and characteristics, indicating that its \textbf{extensibility} is
good. Potential extensions and further aspects of the taxonomy, including its usefulness and ease of use, are discussed in \cref{sec:discussion}.

We visualize the taxonomy (or, strictly speaking, categorized instances) in a compact form using feature vectors with characteristics abbreviated to single-letter codes. This approach has a drawback, as it requires referencing a legend. Additionally, non-applicable characteristics in mutually exclusive dimensions are not visible, which means the design space is not completely shown. However, the compactness of the representation allows LLM components within a common application to be grouped closely, so that an LLM-integrated application can be perceived as a unit without appearing convoluted. This is a significant advantage for our purposes.

\section{Discussion}\label{sec:discussion}

The discussion first focuses on the taxonomy's applicability and ease of use before considering its overall usefulness.

\subsection{Applicability and ease of use}
The taxonomy was effectively applied to LLM-integrated applications based on research papers, source code blog posts, recorded software demonstrations, and developer experiences. 
The analysis of \textsc{LowCode} revealed it to be a prompt definition tool combined with an LLM-based chatbot, which deviates from the strict definition of an LLM-integrated application. Still, the taxonomy provided an effective categorization and led to a clear understanding of the system's architecture. 

Obviously, the ease of categorization depends on the clarity and comprehensiveness of the available information, which varies across analyzed systems.
Analyzing applications of LLMs in novel and uncommon domains can be challenging.
While these papers present inspiring and innovative ideas for LLM integration, such as \textsc{MyCrunchGpt} and \textsc{TruckPlatoon}, they may prioritize explaining the application area and struggle to detail the technical aspects of the LLM integration. A taxonomy for LLM-integrated applications can guide and facilitate the writing process and lead to more standardized and comparable descriptions.

Applying the taxonomy is often more straightforward for research-focused systems. Omitting the complexities required for real-world applications, such as prompt checks and output revisions, their architectures are simpler and easier to describe. A taxonomy can point out such omissions.

A fundamental challenge in applying the taxonomy arises from the inherent versatility of LLMs, which allows to define LLM components serving multiple purposes. This is exemplified by \textsc{Sgp\-Tod PolicyPrompter}, where the prompt is designed to produce a structure with two distinct outcomes (a class label and a chatbot response), and similarly by \textsc{MatrixProduction}, as detailed \cref{sec:cases}.
Drawing an analogy to ``function overloading'' in classical programming, such LLM components can be termed ``overloaded LLM components''.

A taxonomy can handle overloaded LLM components in several ways:
(1) define more dimensions as non-mutually exclusive, (2) label overloaded LLM components as ``overloaded'' without a more detailed categorization, or (3) categorize them by their predominant purpose or output.
While the first approach allows for the most precise categorization, it complicates the taxonomy. Moreover, it will likely result in nearly all characteristics being marked for some LLM components, which is ultimately not helpful. The second approach simplifies categorization but sacrifices much detail. Our taxonomy adopts the third approach, enforcing simplification and abstraction in descriptions of overloaded LLM components while retaining essential detail. The taxonomy can easily be extended to include approach (2) as an additional binary dimension.

\subsection{Usefulness}

The search for instances of LLM-integrated applications uncovered activities across various domains. Substantial research involving LLM integrations, often driven by theoretical interests, is notable in robot task planning \cite{meesGroundingLanguageVisual2023,singhProgPromptGeneratingSituated2023,
wangSafeTaskPlanning2024,liuDELTADecomposedEfficient2024,wangLLM3LargeLanguage2024} and in the TOD field  \cite{hudecekAreLargeLanguage2023,zhangSGPTODBuildingTask2023,bocklischTaskOrientedDialogueInContext2024, caoDiagGPTLLMbasedChatbot2023, thakkarUnifiedApproachScalable2024}. Research exploring LLM potentials from a more practical perspective can be found in novel domains, such as industrial production \cite{XiaAutonomousSystemFlexible2023,kernanfreireHarnessingLargeLanguage2023} and other technical areas \cite{kumarMyCrunchGPTLLMAssisted2023,dezarzaLLMAdaptivePID2023}. 
Furthermore, developers of commercial LLM-based applications are beginning to communicate their efforts and challenges \cite{parninBuildingYourOwn2023,carterAllHardStuff2023}.
 The taxonomy has been applied to example instances from these and additional areas. This demonstrates its potential as a common, unified framework for describing LLM-integrated applications, facilitating the comparison and sharing of development knowledge between researchers and practitioners across various domains.
 
When applying the taxonomy to the example instances, it proved to be effective and useful as an analytical lens. Descriptions of LLM-integrated applications commonly explain background information and details of the application domain in addition to its LLM integration. When used as an analytical lens, the taxonomy quickly directs the analysis towards the aspects of LLM integration, abstracting from the specificities of the domain.

The taxonomy describes how LLM capabilities can be leveraged in software systems, offers inspiration for LLM-based functions, and outlines options for their implementation as follows.
 The \textit{Skills} dimension outlines the range of capabilities an LLM can contribute to an application through a concise set of characteristics, while the \textit{Function} dimension suggests potential uses, further supported by the \textit{Interaction} dimension. The \textit{Output Type} dimension indicates options for encoding the output of an LLM in formats beyond plain text, making it processable by software. The \textit{Output Consumer} dimension illustrates the diverse ways to utilize or act upon LLM output. Thus, the taxonomy, as intended, spans a design space for LLM integrations.

The sampled LLM-integrated applications showcase the creativity of researchers and developers in applying and exploiting the potentials of LLMs, ranging from straightforward solutions (\myeg \textsc{TruckPlatoon}) to highly sophisticated and technically complex ones (\myeg \textsc{AutoDroid}). When using the taxonomy to inspire innovative uses of LLMs, we recommend supplementing it with descriptions of example applications to enhance its illustrativeness.
The characteristics of the \textit{Skills} dimension are derived pragmatically from the investigated example instances. While they do not claim to be exhaustive or deeply rooted in LLM theory or cognitive science, they add relevant details to the categorizations and illustrate design options and potentials for using LLMs as software components.

It emerged as a key insight of this research that, rather than
analyzing an LLM-integrated application in whole, analysis should start with
the identification and description of its distinct LLM components. This is
essential for gaining a clear understanding of how the application utilizes the
capabilities of LLMs. The LLM-integrated application then manifests as a combination of its LLM components. 
As shown in \cref{fig:taxsum}, the visualization effectively displays both the quantity and the variety of LLM components in an LLM-integrated application. 

LLM components interact through prompt chaining, where one LLM component's output feeds into another's input \cite{wuAIChainsTransparent2022b}.
When an LLM-integrated application involves such an interaction, the taxonomy
represents it as an \textit{LLM} characteristic within a \textit{Prompt}
dimension. The taxonomy can capture the variance in these interactions. For
instance, in \textsc{AutoDroid TaskExecutor} and \textsc{LowCode Executing},
the \textit{LLM} characteristic appears in the \textit{Prompt State} dimension,
because their prompt components (knowledge base excerpts and prompt definition,
respectively) are generated by other LLM components in a preparatory stage.
In contrast, the \textit{LLM}
characteristic appears in the \textit{Prompt Task} dimension for \textsc{MatrixProduction Operator}, because its prompt
part is generated individually by the \textsc{MatrixProduction Manager}  almost immediately before use.

Taxonomy dimensions that cover entire LLM-integrated applications may be useful.
Given their complexity, these dimensions should be designed based on a broader
range of examples, which will only become available as more LLM-integrated
applications are developed and their architectures disclosed in the future.
Extensions to the taxonomy could also include dimensions for describing the
structure of prompts in more detail, as well as dimensions addressing
characteristics of the language models used.

\begin{table*}[t!]
\caption{LLM usage in the sample instances. ``Evals'' indicates evaluations of various LLMs.} \label{tab-llms}
\medskip

\begin{tabularx}{\textwidth}{lllX}
\hline
Application       & Used or best LLM       & Evals       & Comments \\
\hline 
\textsc{Honeycomb} & GPT-3.5& yes   & GPT-4 far too slow \\
\textsc{LowCode}     & GPT-3.5-turbo  &       &  \\
\textsc{MyCrunchGpt}       & GPT-3.5    &       & then awaiting the publication of GPT-4    \\
\textsc{MatrixProduction}  &  text-davinci-003      &       &       \\
\textsc{WorkplaceRobot}    & GPT-3  &       &    \\
\textsc{AutoDroid} & GPT-4  & yes   & GPT-4 best for tasks requiring many steps    \\
\textsc{ProgPrompt} & GPT-3   &       & CODEX better, but  access limits prohibitive \\
\textsc{FactoryAssistants} & GPT-3.5&       &    \\
\textsc{SgpTod}    & GPT-3.5& yes   & GPT-3.5 best more often than others combined       \\
\textsc{TruckPlatoon}      & GPT-3.5-turbo  &       &      \\
\textsc{ExcelCopilot}     & N/A      &       & combined LLMs in Copilot for Microsoft 365  \cite{packMicrosoftCopilotMicrosoft2024}  \\
\hline
\end{tabularx}
\end{table*}

\section{Conclusion}\label{sec:conclusion}

This paper investigates the use of LLMs as software components. Its perspective differs from  current software engineering research, which investigates LLMs as tools for software development \cite{fanLargeLanguageModels2023,HouLargeLanguageModels2023} and from research examining LLMs as autonomous agents \cite{chengExploringLargeLanguage2024,wangSurveyLargeLanguage2024,topsakal2023creating,handlerTaxonomyAutonomousLLMPowered2023}. 
This paper defines the concept of an LLM component as a software component that realizes its functionality by invoking an LLM. While LLM components implicitly appear in various works, termed, for example, ``prompters'', ``prompted LLM'', ``prompt module'', or ``module'' \cite{leePromptedLLMsChatbot2023,zhangSGPTODBuildingTask2023,caoDiagGPTLLMbasedChatbot2023,carterAllHardStuff2023}, to our knowledge, this concept has not yet been formalized or systematically investigated.

The main contribution of this study is a taxonomy for the analysis and description of LLM components, extending to LLM-integrated applications by characterizing them as combinations of LLM components. In addition to the dimensions and characteristics of the taxonomy, the study contributes a taxonomy visualization based on feature vectors, which is more compact than the established visualizations such as morphological boxes \cite{szopinskivisualize2020} or radar charts. It represents an LLM-integrated application as one visual entity in a tabular format, with its LLM components displayed as rows.

The taxonomy was constructed using established methods, based on a set of example instances, and evaluated with a new set of example instances. The combined samples exhibit broad variation along the identified dimensions. For some instances, information was not available, necessitating speculative interpretation. However, since the sample is used for identifying options rather than quantitative analysis, this issue and the representativeness of the sample are not primary concerns. The evaluation was conducted by the developer of the taxonomy, consistent with recent related work \cite{handlerTaxonomyAutonomousLLMPowered2023,strobelExploringGenerativeArtificial2024,rittelmeyerMorphologicalBoxAI2023}. Using a new sample for evaluation strengthens the validity of the results.

A further significant contribution of the paper is a systematic overview of a sample of LLM-integrated applications across various industrial and technical domains, illustrating a spectrum of conceptual ideas and implementation options.

As the examples show, LLM components can replace traditionally coded functions in software systems and enable novel use cases. However, practical challenges persist.
Developers report that new software engineering methods are required, \myeg for managing prompts as software assets and for testing and monitoring applications. For instance, the costs of LLM invocations prohibit the extensive automated testing that is standard in software development practice \cite{parninBuildingYourOwn2023, carterAllHardStuff2023}.
 Challenges also arise from the inherent indeterminism and uncontrollability of LLMs. Small variations in prompts can lead to differences in outputs, while automated output processing in LLM-integrated applications requires the output to adhere to a specified format.

Furthermore, the deployment mode of LLMs, whether local (on the same hardware as the application) or remote, managed privately or offered as Language-Models-as-a-Service (LMaaS), has impact on performance and usability. \Cref{tab-llms} gives an overview of the LLMs used in our sample of applications. Where papers report evaluations of multiple LLMs, the table displays the chosen or  best-performing LLM. Although not representative, the table provides some insights. 
LMaaS dominates, likely due to its convenience, but more importantly, due to the superior performance of the provided LLMs. 

Concerns regarding LMaaS include privacy, as sensitive data might be transmitted to the LLM through the prompt \cite{wenEmpoweringLLMUse2023}, and service quality, \myie reliability, availability, and costs. Costs typically depend on the quantity of processed tokens. This quantity also affects latency, which denotes the processing time of an LLM invocation. A further important factor for latency is the size of the LLM, with larger models being slower \cite{carterAllHardStuff2023}.

When building LLM-based applications for real-world use, the reliability and availability of an LMaaS are crucial. Availability depends not only on the technical stability of the service,  but also on factors such as  increased latency during high usage periods or usage restrictions imposed by the provider of an LMaaS, as reported for  \textsc{ProgPrompt} \cite{singhProgPromptGeneratingSituated2023}.  Beyond technical aspects, the reliability of an LMaaS also encompasses its behavior. For instance, providers might modify a model to enhance its security, potentially impacting applications that rely on it.

Despite practical challenges, integrating LLMs into systems has the potential
to  alter the way software is constructed and the
types of systems that can be realized. 
Prompts are central to the functioning of LLM components which pose specific requirements such as strict format adherence. Therefore, an important direction for future research will be prompt engineering specifically tailored for LLM-integrated applications. 

In future work, the taxonomy will be extended to distinguish finer-grained parts of prompts, allowing a more detailed description and comparison of prompts and related experimental results.
Initial studies share results on the format-following behavior of LLMs \cite{xiaFOFOBenchmarkEvaluate2024} as a subtopic of instruction-following \cite{zhouInstructionFollowingEvaluationLarge2023a}, derived with synthetic benchmark data.
It is necessary to complement their results with experiments using data and tasks from real application development projects because, in the early stages of this field, synthetic benchmarks may fail to cover relevant aspects within the wide range of possible options. 
Another crucial research direction involves exploring how LLM characteristics correspond to specific tasks, such as determining the optimal LLM size for intent detection tasks. The taxonomy developed in this study can systematize such experiments and their outcomes. Additionally, it provides a structured framework for delineating design choices in LLM components, making it a valuable addition to future training materials.

\section*{Acknowledgements}
Special thanks to Antonia Weber and Constantin Weber for proofreading and providing insightful and constructive comments.

\bibliographystyle{plainnat} 
\providecommand{\noopsort}[1]{}

\end{document}